\documentclass[12pt, preprint]{aastex}

\begin{document}

\title{DIAGNOSTICS OF NEUTRON STAR AND BLACK HOLE X-RAY BINARIES WITH X-RAY SHOT WIDTHS}

\author{
Hua Feng\altaffilmark{1},
T. P. Li\altaffilmark{2,3}, and
S. N. Zhang\altaffilmark{2,4,3,5}}

\altaffiltext{1}{Department of Engineering Physics and Center for
Astrophysics, Tsinghua University, Beijing 100084, China}

\altaffiltext{2}{Physics Department and Center for Astrophysics,
Tsinghua University, Beijing 100084, China}

\altaffiltext{3}{Laboratory for Particle Astrophysics, Institute
of High Energy Physics, Chinese Academy of Sciences, Beijing
100039, China}

\altaffiltext{4}{Physics Department, University of Alabama in
Huntsville, Huntsville, AL 35899, USA}

\altaffiltext{5}{Space Science Laboratory, NASA Marshall Space
Flight Center, SD50, Huntsville, AL 35812, USA}

\begin{abstract}
A statistic $w$, the differential coefficient of the mean absolute
difference of an observed lightcurve, is proposed for timescale
analysis of shot widths. The shortest width of random shots can be
measured by the position of the lower cut-off in the timescale
spectrum of $w$. We use the statistic to analyze X-ray lightcurves
from a sample of neutron star and black hole binaries and the
results show that the timescale analysis can help us distinguish
between neutron star binaries and black hole binaries. The
analysis can further reveal the structure and dynamics of
accretion disks around black holes.
\end{abstract}

\section{Introduction}
Complex X-ray emission variability at wide timescales has been
observed in accreting stellar mass black hole candidate (BHC) and
weakly magnetized neutron stars (NS) systems \citep[for reviews
see][]{tan95, van95}. On the subsecond timescale, aperiodic
fluctuations are common in these systems, as first discovered by
\citet{oda71} in the prototype BHC Cyg~X-1 in its hard state with
a large amplitude (up to $\sim 40\%$ rms of the mean flux). Many
properties of X-ray rapid variability from Cyg~X-1, e.g., the
Fourier power spectrum, the mean and variance, and the
autocorrelation function of the lightcurve, could be described as
the superposition of individual shots with the characteristic
duration of a few tenths of a second \citep{ter72, oda77, sut78,
nol81, mee84, miy88, miy92, loc91}. The average shot profiles of
Cyg~X-1 have been obtained by superposing many shots by aligning
their peaks in the lightcurves obtained with \textit{Ginga}
\citep{neg94, neg01} and \textit{RXTE} \citep{fen99}.

Studying the short timescale variability of the X-ray emission of
accreting BHC or NS systems is important for understanding the
emitting region and emission process of high energy photons, as
well as the nature of the compact object. Timing studies carried
out in the time domain can make direct measurements of, e.g.,
power, spectral lag, of a stochastic process at high frequencies
(short timescales). \citet{lit01} has developed a technique of
timescale analysis, with which timescale spectra can be derived
directly in the time domain. The timescale analysis of the
correlation coefficient between intensity and hardness ratio of
Cyg~X-1 \citep{lit99} is consistent with the results from
superposed shots \citep{fen99, neg01}. The time domain power
density at a certain timescale can be defined as the differential
of the variation power in a lightcurve with the corresponding time
step \citep{lit01}. The timescale spectra of power density of a
sample of NS and BHC binaries were analyzed and the results
indicate that the characteristic timescale of hard X-ray shots is
$\sim 0.1$~s for BHCs, and much shorter ($\ll 1$~ms) for NSs
\citep{lit02}.

In the present work, we propose a statistic $w$, the differential
coefficient of the mean absolute difference of an observed
lightcurve, to make timescale analysis of shot width. The
timescale spectrum of $w$ can reflect the width distribution of a
stochastic shot process with noise. We introduce the method of $w$
analysis in \S 2. We studied the $w$ spectra of a sample of NS and
BHC binaries and made an energy dependent analysis of the shot
widths for three BHC binaries with RXTE/PCA data; the results are
presented in \S 3. In \S 4 we discuss the possible implications of
our results.

\section{Method}
\subsection{Algorithm}
Let $\{x(j;\delta t)\}$ denote the originally observed lightcurve,
where $x(j;\delta t)$ (cts/s) is the counting rate during the time
interval $[j\delta t, (j+1)\delta t)$ ($j=0,1,2,\ldots$), and
$\delta t$ is the time resolution. To study the variability on a
larger timescale $\Delta t=M\delta t$, we need to construct a
lightcurve with the time step $\Delta t$ from the original time
series by combining its \textit{M} successive bins as
\begin{equation}\label{x}
x_{m}(i;\Delta t) =
\frac{1}{M}\sum_{j=iM+m}^{(i+1)M+m-1}x(j;\delta t) \quad \mbox{cts/s},
\end{equation}
where $i=0,1,\ldots,N-1$ and $m$ ($\in [0,M-1]$) is the discrete
phase of the time series. As the lightcurve $\{x_m(i; \Delta t)\}$
does not include any information about the variation on timescales
shorter than $\Delta t$, it can be used to study the variability
over the region of timescales $\ge \Delta t$. A mean absolute
difference of the observed lightcurve at the timescale $\Delta t$
can be defined as
\begin{equation}\label{W}
W(\Delta
t)=\frac{1}{M}\sum_{m=0}^{M-1}(\frac{1}{N-1}\sum_{i=0}^{N-2}
\frac{|x_m(i+1; \Delta t)-x_m(i; \Delta t)|}{\Delta t})~.
\end{equation}
The quantity $W(\Delta t)$ includes information of variability
with timescales larger than or equal to $\Delta t$. To extract the
information at the timescale $\Delta t$, we can calculate the rate
of change of $W(\Delta t)$
\begin{equation}\label{w}
w(\Delta t)=-\frac{d W(\Delta t)}{d \Delta t}~.
\end{equation}

To detect the signal in a spectrum $w(\Delta t)$ against a noise
background, we need to know the timescale spectrum
$w_{noise}(\Delta t)$ of a time series consisting of only noise.
We assume two independent random variables $X_1$ and $X_2$ follow
the Poisson distribution with the same mean $\overline{X}$ (cts),
then the expectation of their absolute difference value is given
by
\begin{equation}\label{e}
E(|X_{1} - X_{2}|) =
2e^{-2\overline{X}}\sum_{n=0}^{\infty}nI_{n}(2\overline{X})~,
\end{equation}
where $I_{n}$ is the modified Bessel function of the first kind
with the order number $n$. Therefore the mean absolute difference
for the background can be derived as
\begin{equation}\label{Wn}
W_{noise}(\Delta t)=\frac{E(|X_1-X_2|)/\Delta t}{\Delta t}
=\frac{2e^{-2\bar{x}\Delta
t}\sum_{n=0}^{\infty}nI_{n}(2\bar{x}\Delta t)}{\Delta t^{2}}
\end{equation}
with $\bar{x}$ being the average counting rate of the lightcurve,
then
\begin{equation}\label{wn}
w_{noise}(\Delta t)= -\frac{dW_{noise}(\Delta t)}{d\Delta t}~.
\end{equation}
Finally the signal's value of the statistic $w$ can be derived as
\begin{equation}\label{ws}
w_{s}(\Delta t) = w(\Delta t) - w_{noise}(\Delta t)
\end{equation}

Calculating $w_{s}(\Delta t)$ by Eq. (\ref{ws}) at different step
$\Delta t$ for an observed lightcurve  we can get a timescale
spectrum of $w_s$. The differential coefficients in  Eqs.
(\ref{w}) and (\ref{wn}) can be calculated numerically. In
practice, a lightcurve may be divided into $L$ segments,  we can
calculate $w_{s,l}$ for each segment and then obtain the average
value
\begin{equation}\label{mw}
\overline{w}_{s}=\sum_{l=0}^{L-1}w_{s,l}/L
\end{equation}
and its standard error
\begin{equation}\label{stdw}
\sigma(\overline{w}_{s})=\sqrt{\sum_{l=0}^{L-1}(w_{s,l}-\overline{w}_{_s})^{2}/L(L-1)}~.
\end{equation}

\subsection{Simulation}
\begin{figure}
  \plotone{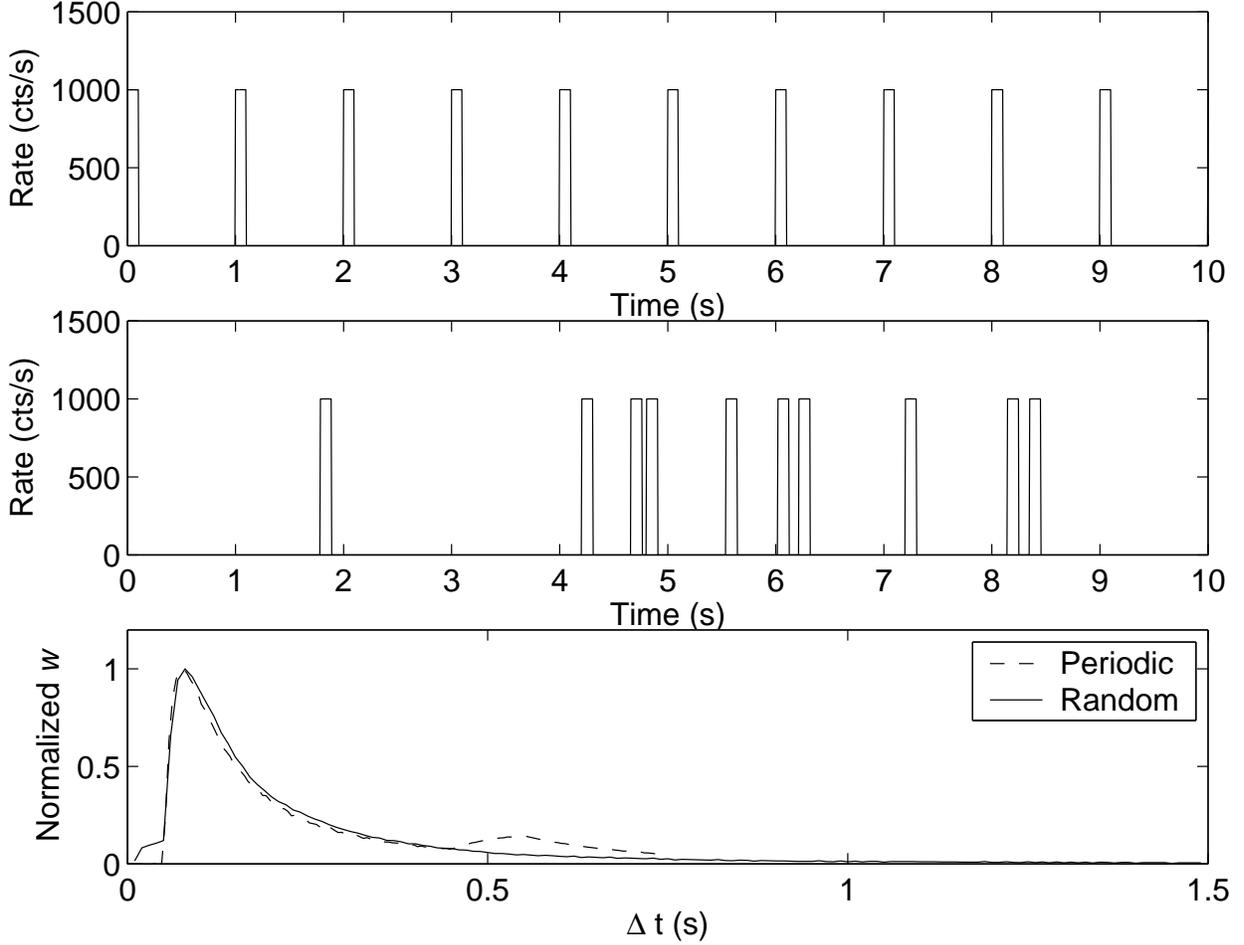}
  \caption{Timescale spectra of $w$ for periodic and random square signals.
  Top: a periodic square signal. The square width $t_w=0.1$ s and the
  period $T=1$ s.
  Middle: a random square signal. The square
  width $t_w=0.1$ s and the interval time between every two
  neighboring squares follows the exponential distribution with a mean of 1 s.
  Bottom:
  \textit{dashed line} - the $w$ spectrum of the periodic signal;
  \textit{solid line} - the $w$ spectrum of the random signal.
  The $w$ spectra reveal the main timescale $t_w$ of both series;
  the periodic process dominated by $T$ in the periodic series is revealed in the $w$
  spectrum at $T/2$; larger amplitude at short timescales of the random series's
  $w$ spectrum indicates the stronger variability at short timescales of the random
  series than the periodic one.}
  \label{pr}
\end{figure}

We make simulation study in order to understand the timescale
spectra of $w$. The $w$ spectra of two time series consisting of
square shots are calculated. The shot width is 0.1~s. One time
series (top panel of \ref{pr}) consists of periodic shots with a
period of 1~s, and in the other time series (middle panel of
\ref{pr}) shots are randomly distributed with shot separations
following an exponential distribution with a mean of 1~s. The time
resolution is 5~ms and the total time length is 1250s with each
segment of 250~s. A peak around the timescale 0.1~s appears
clearly in each spectrum (see the bottom panel of Fig. \ref{pr}).
The other characteristic timescale of the periodic series, 0.5~s,
is also shown in its $w$ spectrum (the dashed line in Fig.
\ref{pr}) with a smaller amplitude. At small timescales, the
random shot series has stronger variability than the periodic one,
indicated by the larger amplitude in the $w$ spectrum.

\begin{figure}
  \centering
  \includegraphics[width=8cm]{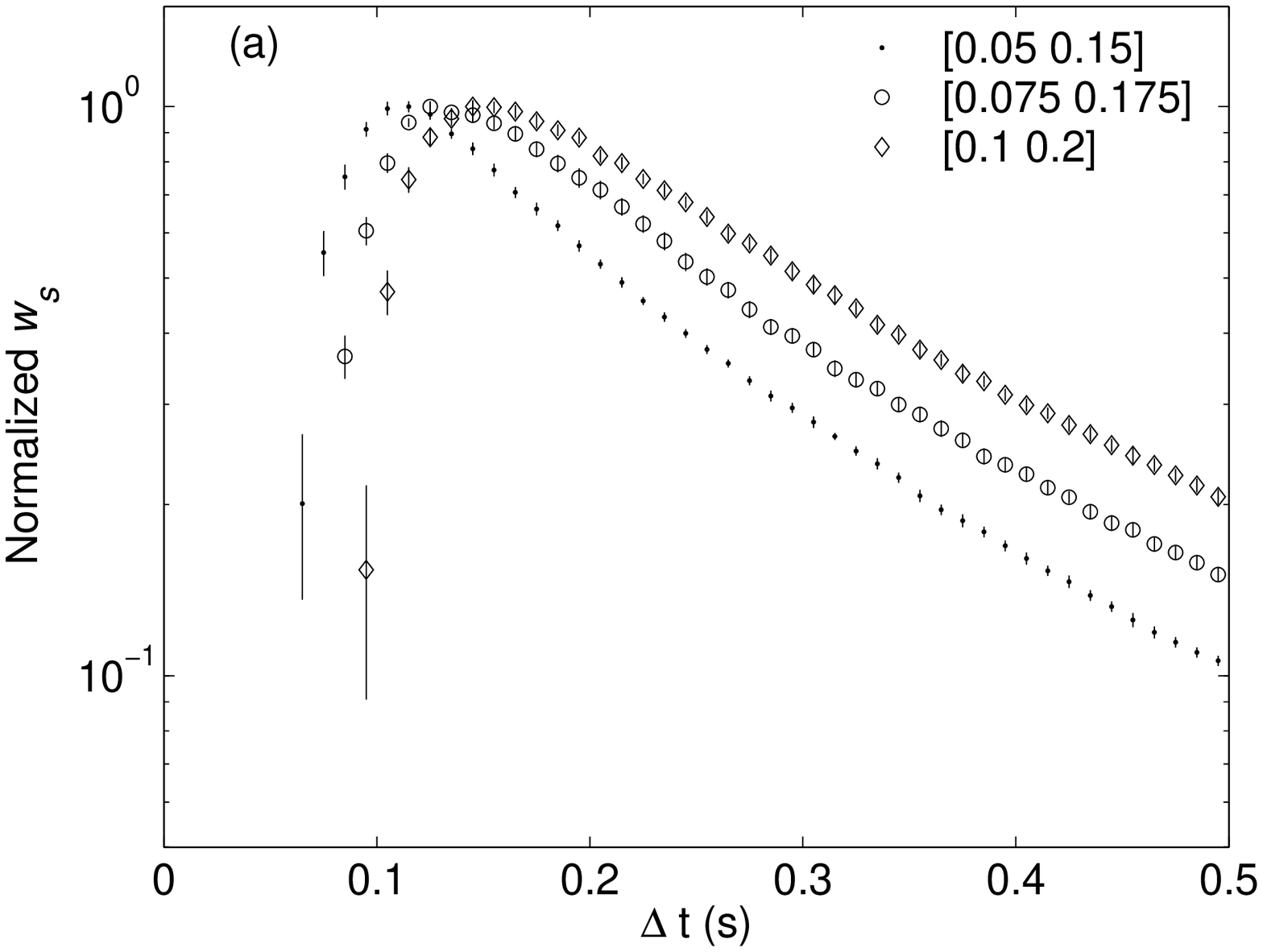}
  \includegraphics[width=8cm]{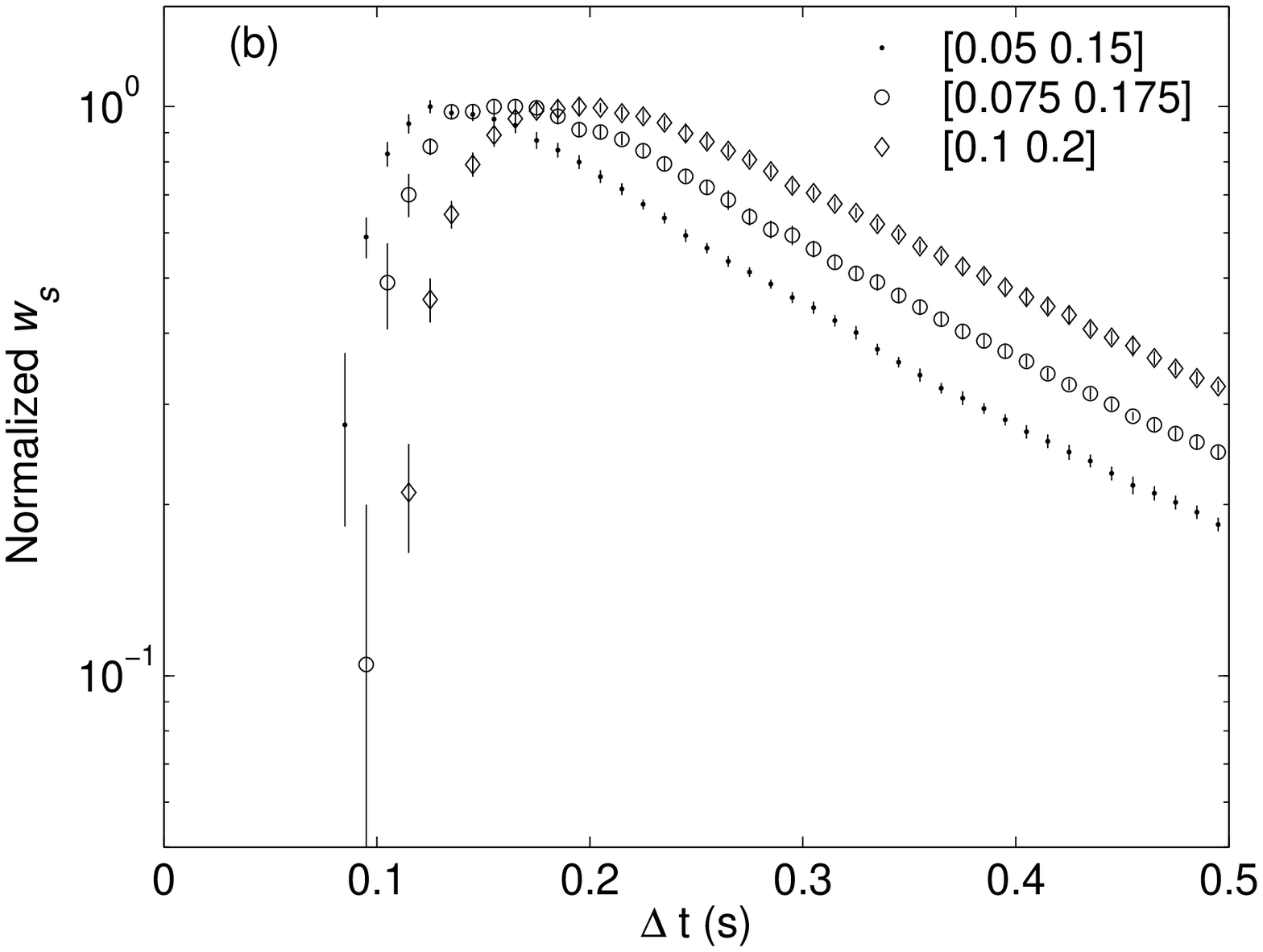}\\
  \includegraphics[width=8cm]{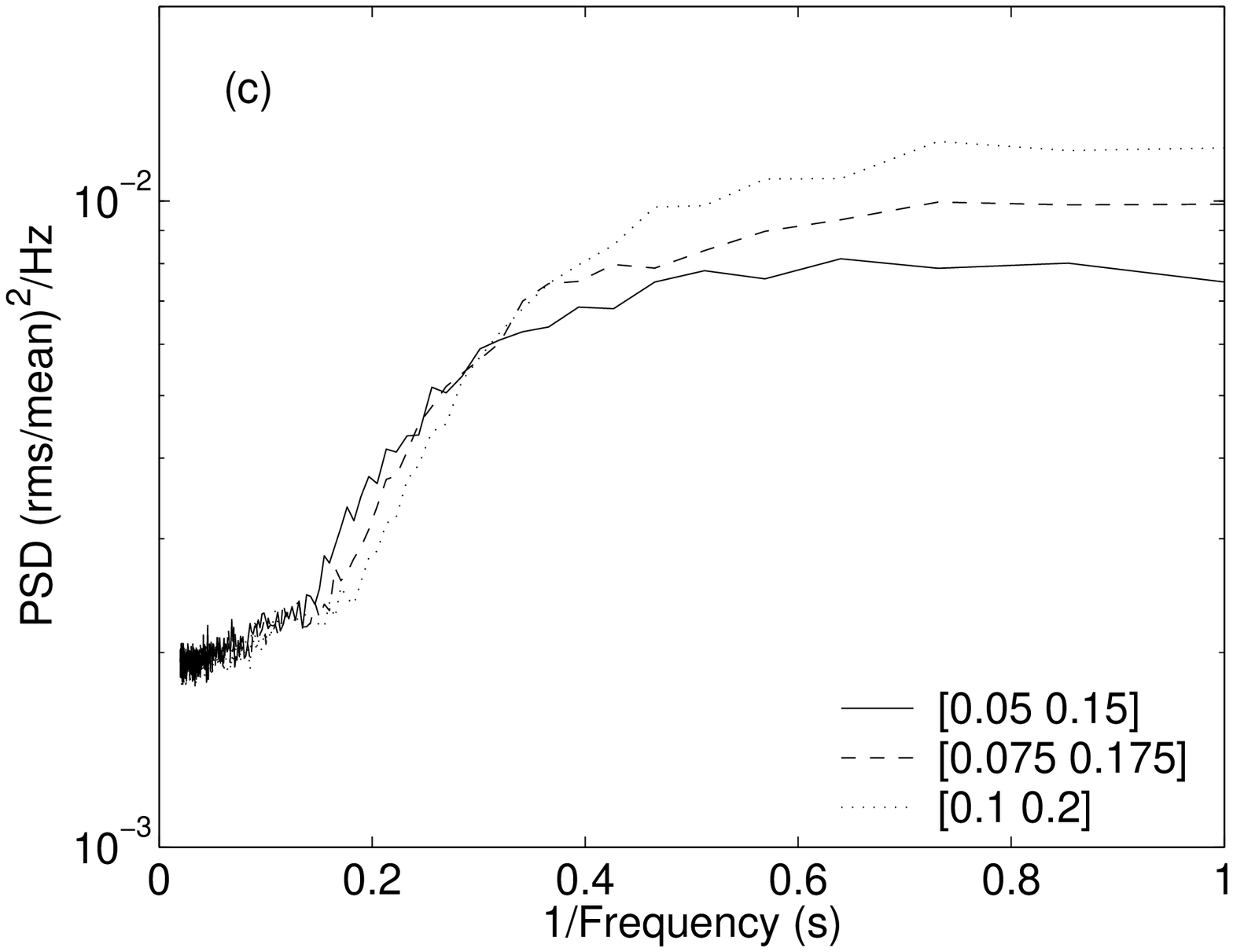}
  \includegraphics[width=8cm]{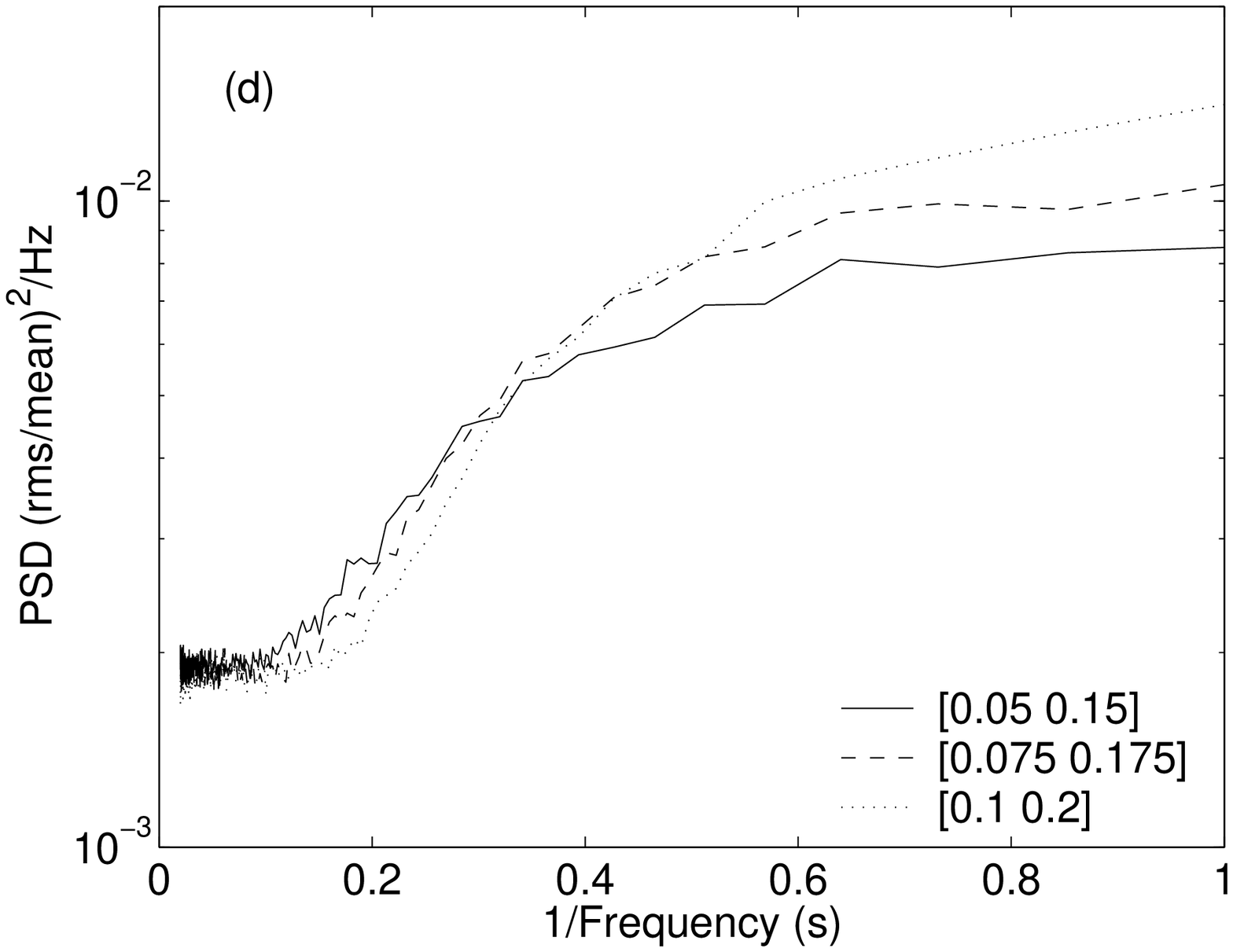}
  \caption{Timescale spectra of $w_s$ for randomly occurring shots. Shots separations are
  exponentially distributed with a mean of 1 s. The peak rate of a shot is random from 0 to
  1000~cts/s above a background of 1000~cts/s. Poisson counting statistics are
  included in the simulations. The legend means the timescale regions in which shot width
  are randomly distributed (width refers to FWHM for Gaussian shots).
  (a) --- $w$ spectra for square shots.
  (b) --- $w$ spectra for Gaussian shots.
  (c) --- PSDs for square shots.
  (d) --- PSDs for Gaussian shots.
  The lower cut-off in each $w$ spectrum reflects the smallest
  shot width of the shot series. By fitting the cut-off edge of the $w$ spectrum,
  the shortest shot width can be measured (see details in \S 2.2). From
  the simulation, we can see $w$ spectrum is better than Fourier
  technique in shot width detection.}
  \label{shots}
\end{figure}

The $w_s$ spectra of a set of simulated lightcurves consisting of
both white noise and signal shots randomly occurring with
different widths and amplitudes are also calculated. The shot
width is randomly sampled from a uniform distribution between
$[\tau_1,\tau_2]$ and its profile has either square or Gaussian
form. The peak amplitude of a shot is drawn from a uniform
distribution between 0 and 1000~cts/s. The shot occurring time is
randomly distributed following an exponential distribution with a
mean of 1~s. The time resolution is 10~ms and the total duration
is 5000~s with each segment of 500~s. A background of 1000~cts/s
is added. Poisson counting statistics are also included in the
simulations. The results, shown in Fig. \ref{shots} (panel (a) for
square shots and (b) for Gaussian shots), indicate that we can use
the position of short timescale cut-off in the $w_s$ spectrum to
measure the shortest width of random shots in an observed
lightcurve with noise. For comparison, we also calculate the
Fourier power spectrum density (PSD) for each shot series. The
results are plotted in panel (c) and (d) of Fig. \ref{shots} for
square and Gaussian shots, respectively. It is clear by comparing
the PSD with the $w_s$ spectra in Fig. \ref{shots} that the $w$
spectrum can be used to detect the shot width more conveniently
and accurately.

To measure the lower cut-off in a $w_s$ spectrum, we fit a
polynomial to the front edge and obtain the root value. The error
of the cut-off timescale is evaluated from the fitting error with
90\% confident level. In this way, the lower cut-off timescales
for the above simulations are $0.0614\pm0.0006$, $0.0743\pm0.0008$
and $0.0893\pm0.0009$ (s) respectively for different lightcurves
in the left panel of fig.~\ref{shots}, and $0.0794\pm0.0007$,
$0.0923\pm0.0009$ and $0.1092\pm0.0010$ (s) for the right panel of
fig.~\ref{shots}. The measured lower cut-off timescale in the $w$
spectrum is very close to the input lower scale of shot widths,
which are 0.05, 0.075 and 0.1 (s) respectively for both square and
gaussian shots. There are larger differences for Gaussian shots
because of the Gaussian FWHM may not equal to $w$ spectrum's shot
width. We therefore conclude that we can detect the lower cut-off
timescale of X-ray shots from lightcurves with the $w$ spectrum.

\begin{figure}
  \plottwo{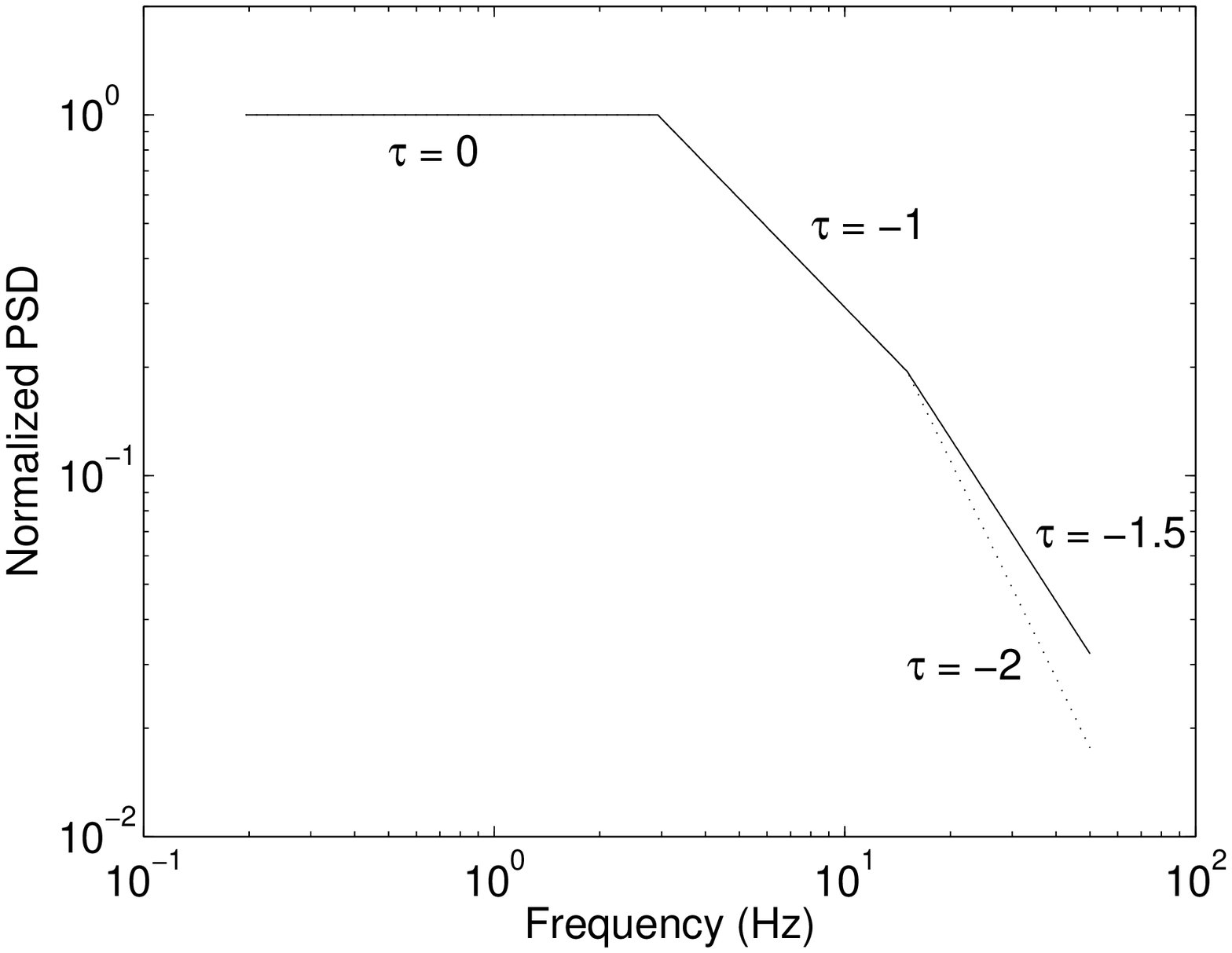}{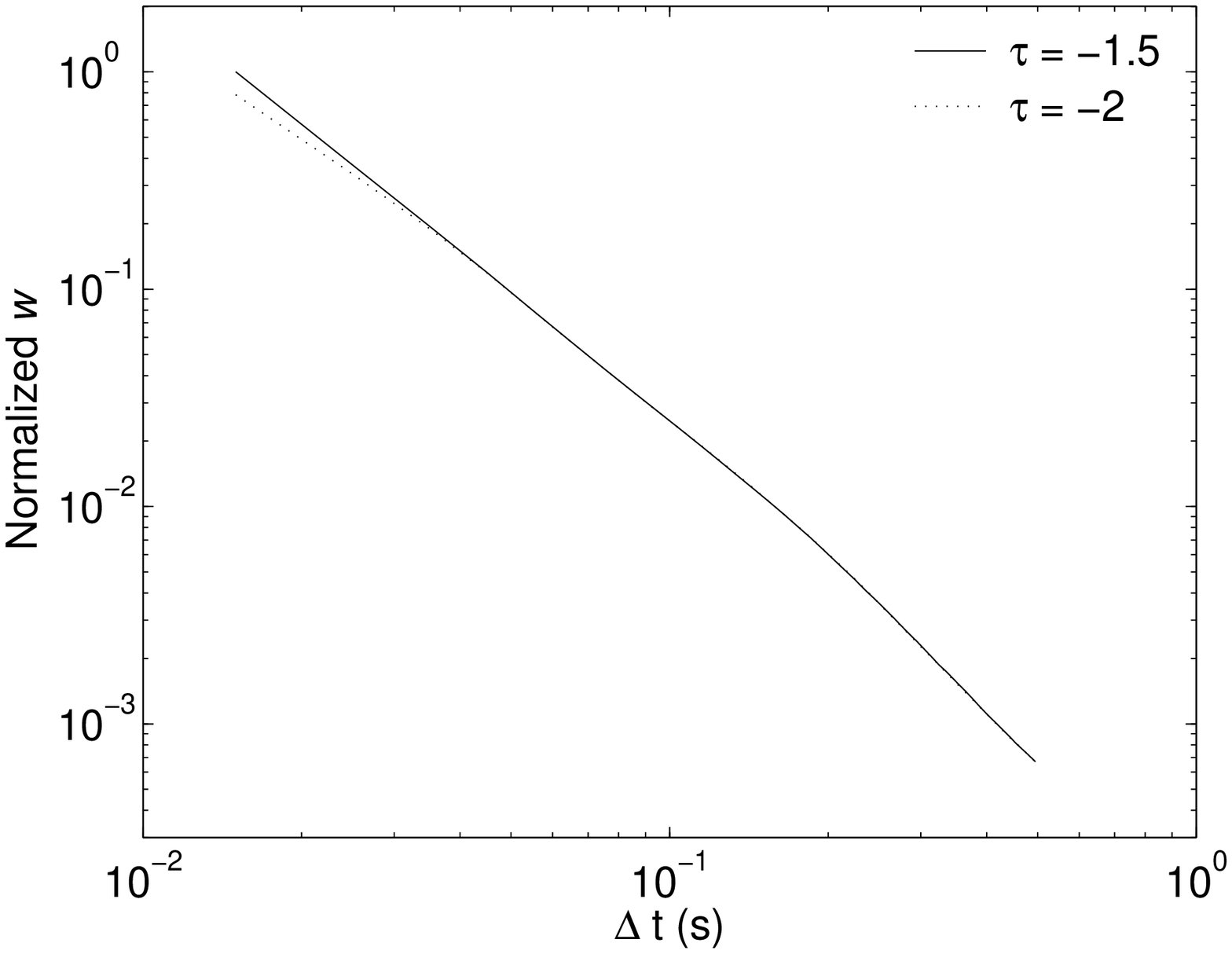}
  \caption{$w$ spectra of lightcurves with different PSD power-law indices. Left: PSDs of
  two lightcurves with power-law index $\tau=0$ when $f < 3 $ Hz,
  $\tau = -1$ when $f \geq 3$ and $< 15$ Hz, and $\tau=-1.5$ or $-2$ when
  $f \geq 15 $ Hz. Right: $w$ spectra of these two lightcurves.
  From the simulation, we can see that the frequency breaks in the PSD do not
  correspond to timescale cut-offs, but manifest as different slopes
  in the $w$ spectra.}
  \label{powerlaw}
\end{figure}

To address the question whether $w$ spectrum may also detect the
frequency breaks often seen in the PSD of the lightcurves many
astrophysical sources, we simulate two lightcurves with different
PSDs \citep[using the inverse FFT method of][]{tim95}. The PSD
consists of power-law forms at different segments: power-law index
$\tau=0$ when $f < 3 $ Hz, $\tau = -1$ when $f \geq 3$ and $< 15$
Hz, and $\tau=-1.5$ or $-2$ when $f \geq 15 $ Hz (left panel of
Fig.~\ref{powerlaw}). The $w$ spectra of these two lightcurves are
presented in the right panel of Fig.~\ref{powerlaw}. We can see
that the frequency breaks in the PSD do not correspond to
timescale cut-offs, but manifest as different slopes in the $w$
spectra.

In sum, the $w$ spectrum is only sensitive to the width
distribution of a shot series, but insensitive to other shot
parameters, such as the shot profile, shot separation distribution
and shot amplitude. In particular, the $w$ spectrum is especially
powerful in detecting the smallest shot width of a shot series.

\section{Data Analysis}
We apply the timescale analysis method to observations of a number
of X-ray binaries with the Proportional Counter Array (PCA) on
board \textit{Rossi X-ray Timing Explorer (RXTE)}. The data are
screened when five PCUs are all on and the original time bin size
$\delta t$ of the lightcurve is binned to 2~ms or 4~ms according
to the data mode, with each segment of 50000 bins (100~s or
200~s). Because this analysis method can tolerate high level of
noise, we did not avoid PCA channels 0-7, which have higher level
of instrumental background. The observation IDs and other details
of used data are listed in Table \ref{data}.

The top panel of Figure \ref{bhns} shows the normalized $w_s$
spectra of five accretion BHCs, GRS~1915+105, GRO~J1655-40,
GX~339-4, XTE~J1550-564, and Cyg~X-1. All spectra show a cut-off
at a timescale about 0.05~s. But for five selected NS binaries
with weak magnetic field, i.e., 4U~1705-440, GS~1826-24,
4U~0614+091, 4U~1608-522, and Cyg~X-2, the $w_s$ spectra have no
obvious cut-off above $10^{-2}$~s (see the bottom panel of Fig.
\ref{bhns}).

The energy dependent analysis is carried out for three BHC
binaries: Cyg~X-1, XTE~J1550-564, and GRO~J1655-40. All sources
are at their hard state. The choice of energy bands for each
observation reflects the different data mode of each observation.
From Figure \ref{ec}, one can see a common trend: below about
10-20 keV the cut-off timescale $\Delta t_c$ decreases with
increasing energy, and above about 10-20 keV $\Delta t_c$
increases with increasing energy. An emission feature at iron
K$\alpha$ line appears clearly in the energy spectra of Cyg~X-1
and XTE~J1550-564, and a broad peak around $\sim$ 6~keV can also
be seen in their $\Delta t_c$ spectra. For GRO~J1655-40, the Fe
reflection component is very weak in the energy spectrum of this
observation, and this broad peak in its $\Delta t_c$ spectra is
also not seen.

\section{Discussion}
\begin{figure}
  \plotone{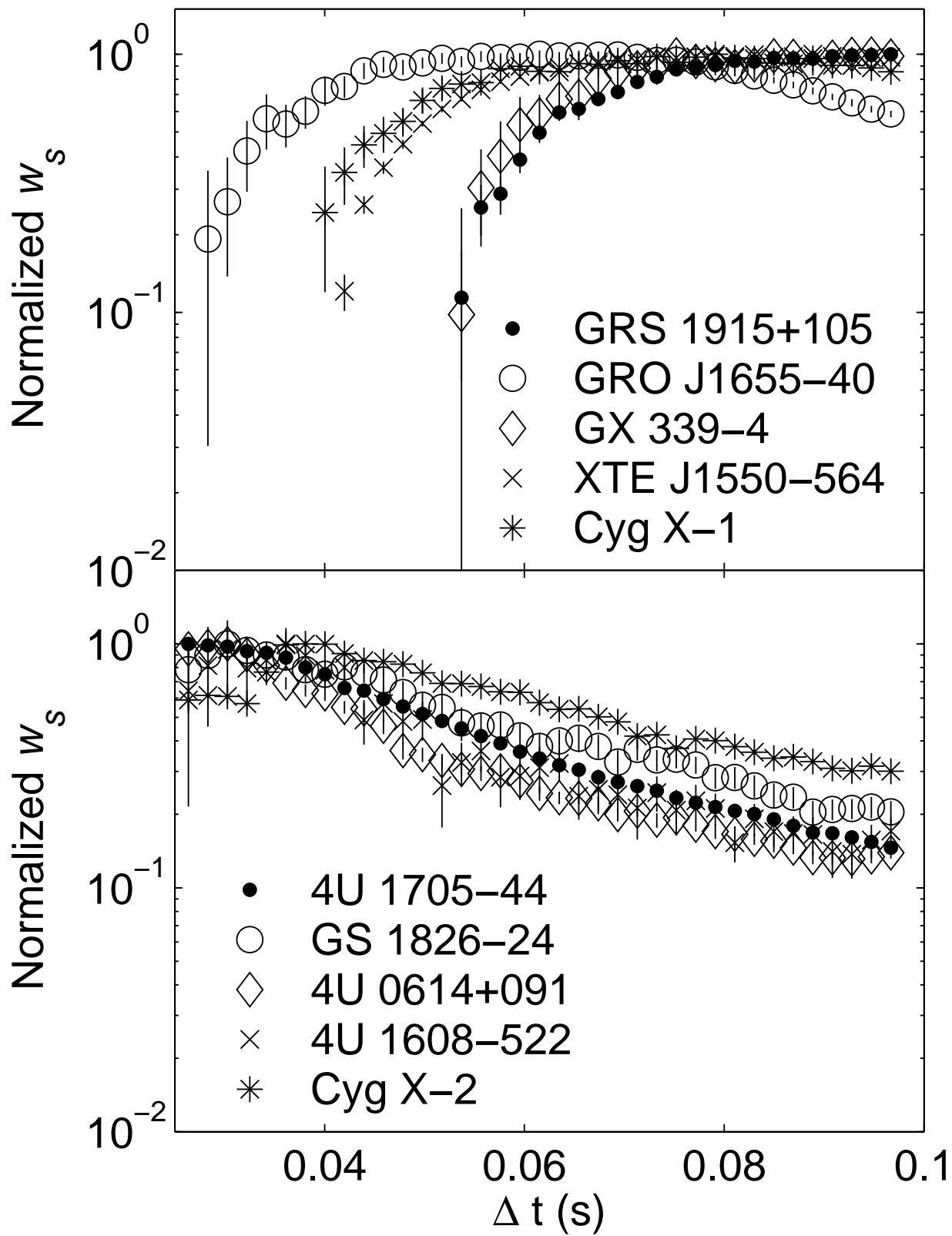}
  \caption{Timescale spectra of $w_s$ for BHC and NS binaries.
  Top: Black hole binary candidates; Bottom: weakly magnetized neutron star binaries.
  All BHC samples show cut-offs below about 0.05 s, contrary to NS binaries.}
  \label{bhns}
\end{figure}

Extracting physical information from lightcurves for
multi-timescale and aperiodic processes is difficult. However, it
is important to study the underlying physics in high-energy
processes around compact objects. The timescale analysis technique
has the freedom of selecting or designing a proper statistic for a
particular purpose. Our results of simulation and data analysis
for accreting X-ray sources show that the proposed statistic $w$
is useful in timescale analysis of shot widths. From their $w$
spectra (the top panel of Fig. \ref{bhns}), we find that the BHCs
have shortest shot width $\sim$ 0.05~s, consistent with that shots
in BHCs have characteristic timescale $\sim$ 0.1~s, which was
previously derived by a timescale analysis of power density
\citep{lit02}. The fact that no short timescale cut-off appears in
the $w$ spectra of NS binaries (the bottom panel of Fig.
\ref{bhns}) is also consistent with what indicated by their
timescale spectra of power density \citep{lit02}, and by Fourier
spectral analysis \citep{sun00} of these systems in the hard
state. The timescale analysis technique of shot width proposed in
this work, like the time domain power spectrum analysis, can help
us distinguish BHCs from NSs. Compared with the timescale analysis
of power density spectrum, the statistic $w$ can measure more
directly and sensitively the lower cut-off of shot timescale, with
which we can study the energy dependence of the cut-off timescale
quantitatively.

\begin{figure}
  \plotone{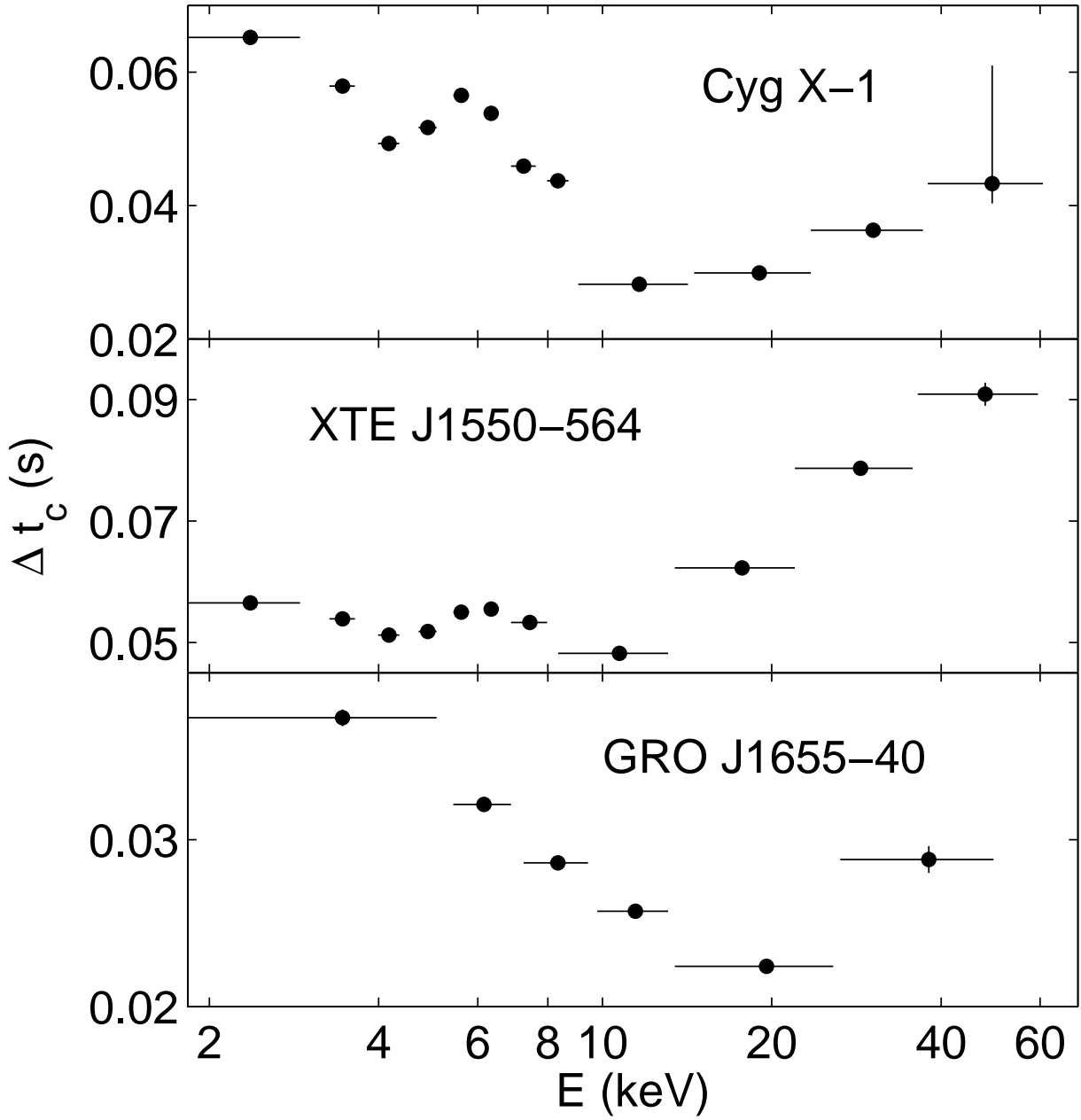}
  \caption{Timescale cut-off vs. energy for three BHCs.}
  \label{ec}
\end{figure}

The variation timescale is usually taken as an indicator of the
spatial dimensions of the underlying physical process. The
remarkable result from our timescale analysis with the statistic
$w$ is the energy dependent timescale cut-off of BHCs (Fig.
\ref{ec}) is hard to explain by a simple correspondence between
the variation timescale and the radius where shots take place. The
monotonic increase of the timescale cut-off with energy in the
region of $E\ge 10$-20~keV is consistent with previous time domain
analysis of variability in Cyg~X-1 and other BHCs
\citep[e.g.,][]{fen99, lit99, mac00, neg01}, which is expected
from the common understanding that hard X-rays come from inverse
Compton scattering of low-energy seed photons by high-energy
electrons in a hot corona \citep[e.g.,][]{ear75, pou96, dov97} or
hot advection-dominated accretion flow (ADAF)
\citep[e.g.,][]{esi98}, where a shot of low energy seed photons
are hardened and broadened by the Comptonization process.

A rough estimate from Figure \ref{ec} can be made to evaluate the
corona size $R_c$ of a BHC. We assume a typical temperature of
$\sim$100 keV for the hot corona at the hard spectral state. The
mean free path of X-rays in corona can be considered as the
corona's spatial dimension $R_c$, because its optical depth is $<$
1 from the energy spectral fitting. We assume the turning point
energy $E_0$ at $w$ spectrum as the energy of input seed shot
photons, and after $k$ times of inverse Compton scattering, the
mean photon energy is increased to $E_k$, the timescale $\Delta
t_{ck}$ of shots at energy $E_k$ can then be located in the
$\Delta t_c$ spectrum. By a simple relationship of $\Delta
t_{ck}=\Delta t_{c0} + k\,R_c/c$, where $\Delta t_{c0}$ is the
turning point timescale in the $\Delta t_c$ spectrum and $c$ is
the light speed, $R_c$ is then estimated. From Figure \ref{ec},
the approximately derived $R_c$ are in order of 30, 140 and 70
$R_s$ (Schwarzschild radius) for Cyg~X-1, XTE~J1550-564 and
GRO~J1655-40, respectively, assuming these BHC masses are 10, 10
and 7 $M_\odot$ \citep{qui02}.

It is obvious that the general trend of timescale cut-off vs.
energy in the energy region of $E\le 10$-20~keV is opposite to
that in higher energy region. We first ignore the broad peak
around 6 keV in the $\Delta t_c$ spectrum and focus on the
decreasing trend of $\Delta t_c$ with increasing energy at $E\le
10$-20~keV. One probability is that shots of different energies
are produced in different regions of the disk: the high energy
shots are located at inner disk and low energy ones are from outer
region. The turning point at $\sim$~10-20~keV may indicate that
the maximum energy of the initial shot photons, which are most
probably produced at the turbulent region of the cold disk joining
the hot corona, can be as high as up to $\sim$ 10-20~keV
\citep{man96,lit99}.

Another possibility to explain the $\Delta t_c$ spectrum at energy
below the turning point is to consider the shots with energies
below the turning point are produced by down-Comptonization, i.e.,
original shots with energy at the turning point are scattered by
cold electron cloud covering the accretion disk \citep{zha00}.
From the top panel of figure \ref{ec} (Cyg~X-1), we can see that
from 11.6 keV (the turning point energy) to 2.4 keV (the lowest
energy) the timescale is increased by 0.037 s. If the mean kinetic
energy of cold electrons is much less than the turning point
energy, about 170 times of scattering are needed for the X-ray
energy increased from the turning point energy to 2.4 keV.
Assuming the relationship between the increase of timescale
$\Delta t$, the number of scattering $N$ and mean free path $l$
under the optically thick condition, is described by $\Delta
t=lN/c$, the derived mean free path $l$ is $6.5\times 10^6$ cm
(2.2 $R_s$ of Cyg~X-1).

\begin{figure}
  \plotone{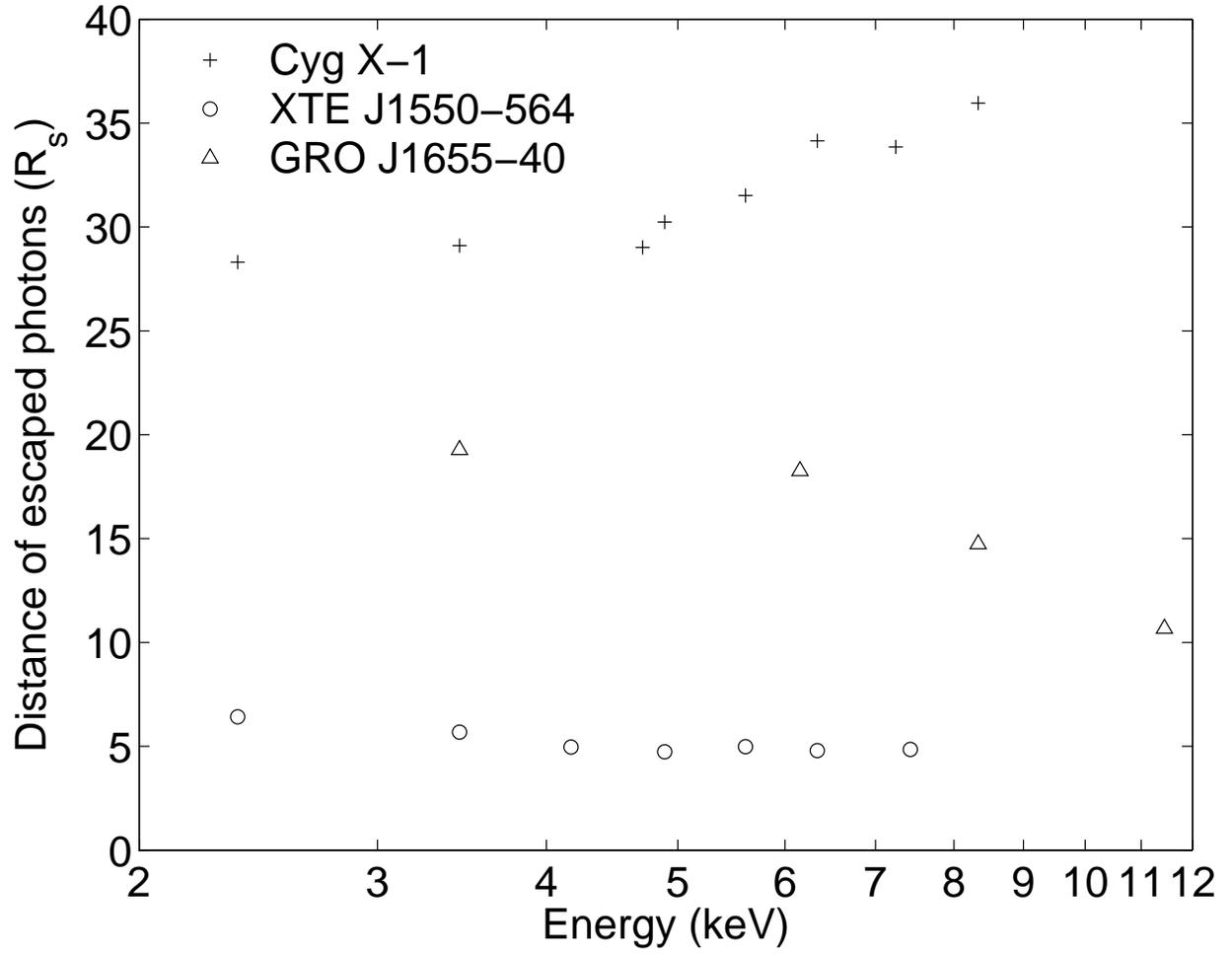}
  \caption{Transportation distance of photons from
  the seed shots location to the surface of the cold electron cloud.}
  \label{ed}
\end{figure}

Under the above assumption that an optically thick cold electron
cloud covers the initial shots (with energy at the turning point
of the $\Delta t_c$ spectrum) located near the innermost region of
the disk, we can evaluate the geometry of the cold electron cloud,
from the shape of the the $\Delta t_c$ spectrum (top panel of fig.
\ref{ec} for Cyg~X-1). The distance between the location where the
down-scattered photons escape to the seed shot photon region can
be described as $D=\tau l$, where $\tau$ is the optical depth and
equal to $\sqrt{N}$ in an optically thick system. We find that
$\tau$ increases and $l$ decreases when the photon energy
decreases, and consequently $D\approx 30R_s$ of $D$ is obtained,
almost independent of the escaped photon energy. Similarly we also
applied the above procedure to the data on XTE~J1550-564 and
GRO~J1655-40. In Figure \ref{ed} we show the relationship between
$D$ and the escaped photon energy.

Our calculation for Cyg~X-1 also indicates that if the mean
temperature of electrons is larger than 0.03 keV, the above
down-Comptonization process cannot decrease the average photon
energy from the turning point to below 2.4 keV. For example, if
the electron temperature is 0.03 keV, after about 1000 Compton
scattering the average photon energy saturates at 2.4 keV, down
from the seed shot photon energy of 11.6 keV. So if the above
assumed down-Comptonization scenario is true for the $\Delta t_c$
spectrum below the turning point, we predict that a sharp increase
of timescale in the $\Delta t_c$ spectrum below the saturation
energy, which is uniquely related to the temperature of the cold
electron cloud.

Finally we examine the broad peak in the $\Delta t_c$ spectrum
around $\sim$ 6 keV for Cyg~X-1 and XTE~J1550-564. It is most
probably produced by the disk reflection component: reprocessed
iron K$\alpha$ fluorescent photons in the cold disk irradiated by
hard X-rays with energy above the $\sim$7keV iron K-edge will
further broaden the shots. This interpretation is also consistent
with the explanation of time delays and the behavior of the iron
line \citep{gil00,pou02} with frequency-resolved energy spectra
\citep{rev99}. Besides producing the fluorescence iron K$\alpha$
emission, the refection also has a strong component at energies
above 10 keV, which may contribute to the increase of shot width
above the tuning point in addition to the up-Comptonization
process we discussed above.

\acknowledgments We are very grateful to the expert and insightful
comments by the anonymous referees; panels (c) and (d) in figure 2
and the whole figure 3 are added in responding to the comments, as
well as several clarifications in the manuscript. This study is
supported in part by the Special Funds for Major State Basic
Research Projects and by the National Natural Science Foundation
of China under project No. 10233030. SNZ also acknowledges
supports by NASA's Marshall Space Flight Center and through NASA's
Long Term Space Astrophysics Program.

\clearpage

\begin{deluxetable}{cllcc}
\tablecaption{RXTE/PCA observations Used} \tablewidth{0pt}
\tablehead{ \colhead{} & \colhead{Object} & \colhead{OBS ID} &
\colhead{Band (keV)} & \colhead{Used at}}

\startdata
 & 4U 1705-440 & 20073-04-01-00 & 3-20 & Fig. \ref{bhns}\\
 & GS 1826-24  & 30054-04-01-00 & 3-20 & Fig. \ref{bhns}\\
Neutron Star & 4U 0614+091 & 30054-01-01-01 & 3-7  & Fig. \ref{bhns}\\
X-ray Binaries & 4U 1608-522 & 30062-01-01-04 & 3-7  & Fig. \ref{bhns}\\
 & Cyg X-2     & 30418-01-01-00 & 2-5  & Fig. \ref{bhns}\\

 \tableline

  & Cyg X-1     & 10236-01-01-03 & 2-67 & Fig. \ref{bhns} \\
 & Cyg X-1     & 40100-01-01-00 & 2-61 & Fig. \ref{ec}\\
& GRS 1915+105& 20402-01-05-00 & 2-5  & Fig. \ref{bhns}\\
Black Hole  & GRO J1655-40& 20402-02-25-00 & 2-5  & Fig. \ref{bhns}\\
X-ray Binaries & GX 399-4    & 20181-01-01-00 & 7-22 & Fig. \ref{bhns}\\
 & XTE J1550-564&30191-01-14-00 & 2-7  & Fig. \ref{bhns}\\
 & XTE J1550-564 & 30191-01-17-00 & 2-60 & Fig. \ref{ec}\\
 & GRO J1655-40 & 10255-01-04-00 & 2-50 & Fig. \ref{ec}\\

\enddata

\label{data}
\end{deluxetable}

\end{document}